\documentclass[pdflatex,sn-mathphys-num]{sn-jnl}

\usepackage{graphicx}%
\usepackage{multirow}%
\usepackage{amsmath,amssymb,amsfonts}%
\usepackage{amsthm}%
\usepackage{mathrsfs}%
\usepackage[title]{appendix}%
\usepackage{xcolor}%
\usepackage{textcomp}%
\usepackage{manyfoot}%
\usepackage{booktabs}%
\usepackage{algorithm}%
\usepackage{algorithmicx}%
\usepackage{algpseudocode}%
\usepackage{listings}%

\theoremstyle{thmstyleone}%
\theoremstyle{thmstyletwo}%

\theoremstyle{thmstylethree}%

\begin{document}

\title[]{Seeing through water: diffuse image-based depth measurements in three-dimensional dam-break flows}


\author*[1]{\fnm{Elia} \sur{Buono}}\email{elia.buono@polito.it}

\author[1]{\fnm{Roberto} \sur{Bosio}}
\author[1]{\fnm{Andrea} \sur{Cagninei}}

\author[1]{\fnm{Davide} \sur{Poggi}}

\affil*[1]{\orgdiv{Department of Environment,
Land and Infrastructure Engineering}, \orgname{Polytechnic of Turin}, \orgaddress{\street{Corso Duca degli Abruzzi}, \city{Torino}, \postcode{10129}, \state{Italia}}}




\abstract{In this work we present a dedicated experimental facility and an image-based method for measuring water depth in a radially spreading dam-break wave propagating over a uniformly illuminated horizontal plane. The facility consists of a prismatic reservoir containing a known volume of water dyed with a soluble colorant and equipped with a removable vertical breach whose geometry can be varied, and a 6.4 m × 3.4 m plane that can be inclined from 0° to 30°. The plane is enclosed within a light box providing highly uniform illumination through an array of 60 LED floodlights. Wave propagation is captured by two scientific CMOS cameras mounted on the ceiling of the light box, which record the spatial and temporal evolution of the dye-induced color intensity associated with the advancing water layer.

Preliminary dry calibration tests were conducted to assess the spectral compatibility between the broadband white-LED emission, the CMOS sensor sensitivity, and the absorption properties of several dyes at different concentrations. This analysis identified the dye providing the highest attenuation within the effective spectral band of the imaging system, ensuring sensitivity to very small optical path lengths. Under these conditions, a green mixture exhibited the most favourable attenuation behaviour. Based on this characterization, a bi-exponential model is introduced to relate the normalized gray level to the optical path length.

A series of dam-break experiments with five initial reservoir levels was performed to assess statistical repeatability. The high consistency observed across the repeated tests confirms the robustness and reliability of the measurement procedure. The validity of the reconstructed depth fields is further supported by independent estimates of the water volume released from the reservoir, obtained from an array of ultrasonic level sensors and from a calibrated analytical emptying model. Together, these comparisons confirm the reliability and accuracy of the proposed methodology.} 


\maketitle

\section{Introduction}\label{Intro}
Dam-break flows represent a critical subject in hydraulic engineering and risk assessment, as they may generate sudden and potentially destructive flooding events \cite{hunt1982asymptotic,baecher1980risk,walder1997methods,wahl2004uncertainty,chanson2006tsunami,thompson2008role}. Despite the inherently three-dimensional nature of real dam failures—often involving complex breach formation processes and strongly spatially variable flow features \cite{macdonald1984breaching,Ajayi2008,soares2007experimental,soares2008dam}—most experimental and numerical studies have traditionally focused on two-dimensional dam-break configurations. These are typically reproduced in channel-like geometries by the instantaneous removal of a gate spanning the full channel width \cite{lauber1998experimentsH,bell1992experimental,aureli2023review}. This emphasis stems from the relative analytical tractability of the two-dimensional problem, its numerical simplicity, and the practical ease of performing such experiments \cite{ritter1892fortpflanzung,whitham1955effects,dressler1952hydraulic,lauber1998experimentsH,lauber1998experimentsS,chanson2009application}. 

Many large dams discharge into deeply incised valleys, where the valley confinement imposes a dominant flow direction. In such settings, despite the inherently three-dimensional nature of real dam-break processes, two-dimensional modelling and laboratory configurations can still provide valuable insight into the formation and propagation of the flood wave. 

In contrast, a substantial number of small- to medium-size reservoirs—such as hillside irrigation basins, artificial ponds for snowmaking, and fire-fighting water storage facilities—release water onto gently sloping or broadly planar terrains rather than into channelized riverbeds. These configurations arguably represent the dam-break scenario with the strongest inherently three-dimensional character, as the resulting surge can spread freely in multiple directions across an unconfined and mildly inclined surface. 

Despite the widespread occurrence of such reservoirs, truly three-dimensional dam-break flows have received remarkably limited experimental attention, with only a few tens of laboratory investigations reported over more than a century \cite{aureli2023review}. Moreover, many of these studies involve lateral contractions or obstacle interactions that induce secondary effects—such as sidewall reflections and wave refractions—thereby limiting their relevance to genuinely unconfined spreading waves \cite{fraccarollo1995experimental,eaket2005use,aureli2008dam,aureli2015experimental,elkholy2016experimental}. Consequently, the state of the art still lacks a systematic experimental characterization of long-term dam-break propagation over wide, effectively two-dimensional surfaces, a configuration requiring flooding areas much broader than the breach width.

The diffuse and accurate measurement of the free surface in three-dimensional, rapidly varying flows remains a major challenge in experimental fluid mechanics. Reliable space–time reconstructions of free-surface elevation are essential not only for the physical interpretation of such phenomena and for the validation of numerical models, including shallow-water solvers, two-phase VOF (Volume Of Fluid) schemes and SPH (Smoothed-Particle Hydrodynamics) methods, but also for hazard  evaluation and risk assessments, which depend critically on estimates of water depth and the associated  flow velocity. dam-break wave propagation represents a paradigmatic example of these challenges, as it combines strong unsteadiness, rapid free-surface deformation and pronounced three-dimensional effects. In this context, high-resolution measurements of water depth are particularly important, especially for fully three-dimensional configurations. A recent review of laboratory dam-break experiments confirms that detailed, spatially resolved free-surface measurements still constitute the primary benchmark for the validation of numerical models \cite{aureli2023review}. 

This need becomes even more apparent when considering advanced numerical approaches. Recent studies have demonstrated the potential of fully three-dimensional lattice Boltzmann and discrete Boltzmann formulations to reproduce dam-break flows, including cases with obstacles or idealized urban layouts, and to improve the prediction of velocity fields compared to depth-integrated shallow-water models \cite{la2020discrete, miliani2021dam}. At the same time, Navier–Stokes solvers coupled with VoF or CLSVoF interface tracking have been shown to be sensitive to grid resolution, turbulence modeling and interface treatment, particularly in highly unsteady regimes involving wave breaking and air–water interaction \cite{bhushan2021assessment}. Under these conditions, high-quality experimental measurements of free-surface evolution remain indispensable to discriminate among modeling assumptions and to provide robust benchmarks for both reduced-complexity and fully three-dimensional numerical solvers. 

The limitations of traditional free-surface measurement devices are well documented \cite{aureli2011image}. Among the most common non-intrusive techniques, ultrasonic level sensors provide precise and reliable point measurements, but they are prone to signal loss when the free surface becomes steep or highly aerated. Pressure transducers, on the other hand, are intrusive and generally unreliable in accelerating flows, where the hydrostatic pressure assumption breaks down. In both cases, the pointwise nature of the measurement requires either a large number of sensors or multiple experimental runs to obtain a sufficiently detailed description of the free-surface evolution. 

These limitations have motivated a growing interest in optical methods capable of providing spatially distributed information. In recent years, LiDAR (Light Detection and Ranging) systems have become increasingly consolidated, as they can reconstruct complex free-surface geometries and acquire up to several hundred profiles per second. However, their scanning acquisition pattern prevents the instantaneous capture of the entire surface, which limits their applicability to rapidly varying phenomena such as dam-break waves \cite{rak2023review,oosterlo2021calibration}. In coastal and riverine environments, several studies have combined laser scanners and video imaging to measure run-up, overtopping, and average free-surface profiles, achieving good accuracy but still facing a trade-off between area coverage and temporal simultaneity \cite{oosterlo2021calibration,vousdoukas2014role}. These challenges have encouraged the adoption of optical imaging techniques that offer spatially distributed and nearly instantaneous measurements of the free surface.

Optical imaging techniques have made significant advances in recent years. Free-Surface Synthetic Schlieren (FS-SS) \cite{moisy2009synthetic} reconstructs instantaneous surface topography by analysing the refraction of a background pattern through image correlation, and recent developments have improved its spatial resolution and algorithmic robustness, broadening its range of applications \cite{li2021single,mermelstein2025modified}. Other approaches, such as stereo photogrammetry/Structure-from-Motion and laser fringe or strip projection, can accurately map aerated or strongly deformed free surfaces when supported by suitable colouring or optical treatments \cite{fleming2019application,rifai2020continuous}. Low-cost depth sensors (e.g. Kinect or Azure Kinect) have also been tested for laboratory measurements of elevation fields, showing promising performance but exhibiting sensitivity to reflections and turbidity that often requires careful calibration or the use of dispersing agents \cite{kim2023uncertainty}. 

Within this context, a particularly simple and effective solution for rapidly varying free surfaces is based on light absorption in transmission. In this approach, the water is dyed with a colourant, the bottom surface is backlit, and a local calibration is used to relate image luminance (or individual RGB channels) to water depth. A single camera can therefore provide spatially distributed free-surface measurements without interfering with the flow, and with an accuracy comparable to that of ultrasonic probes. This methodology was first developed and validated by \citet{aureli2011image}, who demonstrated that spatially distributed calibration enable reliable measurements of dam-break waves, including a discussion of the influence of surface slope and the effect of RAW versus JPEG image formats. Since then, advances in imaging hardware have further strengthened the potential of absorption-based techniques. In the last decade, imaging systems have benefited from high-uniformity LED illumination (which minimizes flicker), CMOS sensors with higher dynamic range and frame rates, and—most notably—access to NIR (near-infrared) bands where water absorbs strongly around 1.44–1.45 µm. Operating in this spectral region can reduce or even eliminate the need for dyes, while maintaining a monotonic and stable relationship between transmitted intensity and water thickness, even under varying temperature conditions. Recent studies in applied optics have indeed exploited the around 1.44–1.45 µm band to achieve robust measurements of the thickness of aqueous films and liquid veils \cite{lubnow2023water}.

In this work we present a new state-of-the-art large-scale facility specifically conceived to investigate three-dimensional dam-break flows spreading over a wide horizontal plane. The system consists of a tiltable reservoir–plane configuration, a 6.4 m × 3.4 m composite floor and a downstream recirculation loop, all enclosed within a light box equipped with highly uniform LED illumination and overhead scientific CMOS cameras. Building on the transmitted-light approach of \citet{aureli2011image}, we develop an image-based water-depth measurement technique tailored to this LED-illuminated, large-area setup. 

The much larger propagation plane considered here cannot be backlit with a conventional panel, as achieving uniform retro-illumination over several square metres is impractical. In addition, uncontrolled ambient light—either natural or artificial—would compromise the stability of the light intensity during the experiments. To overcome these limitations, we adopt an original solution in which the entire experimental domain is enclosed within a large light box designed to provide homogeneous diffuse illumination. The luminous flux and arrangement of sixty LED sources from MBT model M12-D01 (Hangzhou Moonlight Box Technology, Hangzhou. Zhejiang, China) were mounted with optimal orientation to avoid direct illumination of the free surface, thereby eliminating specular reflections and ensuring a uniform radiance field across the measurement plane. This concept is inspired by the light boxes used in professional photography, where diffuse lighting suppresses shadows and glare; in our case, the same principle enables controlled, repeatable and spatially uniform illumination over a significantly larger area. 

Finally, we apply this methodology in a preliminary campaign of 15 dam-break experiments combining three initial reservoir levels $H_o$ with five repeated runs. From the images we reconstruct depth fields $h(x, y, t)$, compute ensemble mean and variability maps, and estimate the total wave volume, which is compared with a simplified emptying model and with independent ultrasonic measurements to provide a robust validation of the method in rapidly varying three-dimensional flows.

\section{Experimental facility}
\begin{figure*}[ht!]
\noindent\includegraphics[width=31pc]{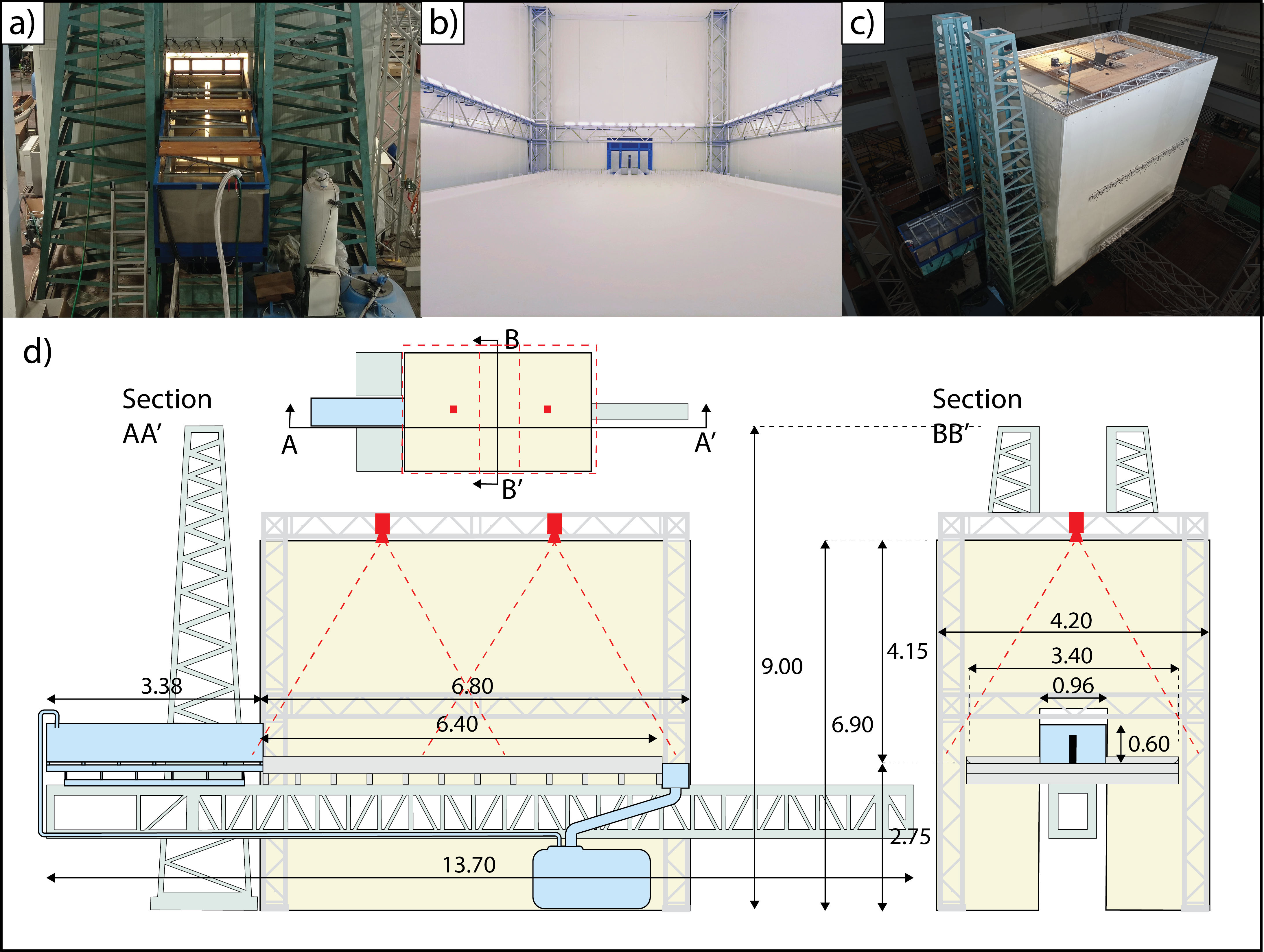}
\caption{The experimental setup for three-dimensional dam-break problem. Panel a) picture of the back side with the reservoir, b) picture from the inside on the light box from the downstream end of the plane, c) picture of the whole facility from the top-upstream corner and d) longitudinal (on the left) and transverse (on the right) sections of the whole experimental facility with the most meaningful dimensions and cameras location.}
\label{Fig:Setup_2D_1}  
\end{figure*}

The study of three-dimensional dam-break flows requires experimental facilities capable of accommodating the inherent complexity of these phenomena. To this end, a state-of-the-art large-scale facility was designed and constructed to provide precise measurements and a high degree of flexibility in the experimental conditions. The setup consists of several interconnected components: a support structure, a prismatic upstream reservoir, a rectangular breach, an instrumented plane floor, a downstream manifold, a recirculating tank, and an enclosing light box, as shown in Figure~\ref{Fig:Setup_2D_1}. The following sections describe each component of the experimental system in detail. 

The support structure forms the backbone of the facility and consists of two steel towers hinged to a $13.7 \, \mathrm{m}$ long lattice beam. This beam provides rigid support for the upstream reservoir, the plane floor and the downstream manifold, allowing these components to be tilted simultaneously. The upstream reservoir is a prismatic stainless-steel tank with internal dimensions of $L_t = 3.38 \, \mathrm{m}$ in length, $W_t = 0.96 \, \mathrm{m}$ in width, and $H_c = 0.6 \, \mathrm{m}$ in high. Its downstream wall incorporates a rigid frame that hosts a removable stainless-steel plate containing a precisely machined breach of $b = 0.05 \, \mathrm{m}$ in width and $H_b = 0.45 \, \mathrm{m}$ in height. This wall is adjustable, ensuring that the breach remains vertical for different imposed longitudinal slopes. 

The plane floor is a composite structure measuring $3.4 \, \mathrm{m}$ in width and $6.4 \, \mathrm{m}$ in length. It consists of several layers: twelve crosswise aluminum beams ($90 \, \mathrm{mm}  \times  \, 180 \mathrm{mm} \times 3.4 \, \mathrm{m}$) that provide structural support; high-rigidity aluminum panels $50 \, \mathrm{mm}$ thick; a $100 \, \mathrm{mm}$-thick lightened concrete layer; and a $2.6 \, \mathrm{mm}$-thick PVC top surface. The aluminum panels furnish a flat and rigid base, while the concrete layer offers a sacrificial surface that can be drilled, modified, or locally damaged as needed, since it can be readily replaced. This layer makes it possible to anchor auxiliary objects (e.g. roughness elements) or to create recesses to house embedded instrumentation such as load cells, without compromising the underlying aluminum structure. The PVC top layer provides a durable, waterproof, low-wettable, and highly reflective surface.

The edges of the floor are bounded by $100 \, \mathrm{mm}$ high banks on the upstream and lateral sides, while the downstream end connects seamlessly to the downstream manifold. This manifold, measuring $0.38 \, \mathrm{m} \times 0.4 \, \mathrm{m} \times 3.4 \, \mathrm{m}$, collects and channels water into a $1.5 \, \mathrm{m}^3$ recirculating tank positioned beneath the floor. Together, these components form a closed-loop system that enables recycling of experimental water.

The plane floor is enclosed within a light box measuring  $4.3 \, \mathrm{m}$ in width, $ 6.8 \, \mathrm{m}$ in length, $6 \, \mathrm{m}$ in height. The light box is built from an   aluminum truss frame and clad with highly reflective white panels, which  enhance illumination uniformity and ensure optimal, repeatable imaging conditions. Sixty LED lamps, with a total electrical power of   $3.6 \, \mathrm{kW}$, are mounted one meter above the plane floor and tilted  $30^\circ$ upward. This configuration prevents direct illumination of the water surface and promotes panel-reflected diffuse light, thereby achieving  homogeneous illumination across the floor and avoiding reflections from the water surface. 

To simulate the dam collapse, the breach in the reservoir's wall is initially sealed with waterproof tape (Flex Tape, Flex Seal products, Weston, Florida, USA) applied to its dry side. This tape is then removed by means of a falling-weight mechanism, which achieves an opening time of approximately $0.05 \, \mathrm{s}$. This rapid breach opening broadly satisfies the standard  criteria for instantaneous dam removal, namely that the opening time be smaller than $\sqrt{2 H_o / g}$ \cite{lauber1998experimentsH}. 
    
\begin{figure*}[ht!]
\noindent\includegraphics[width=31pc]{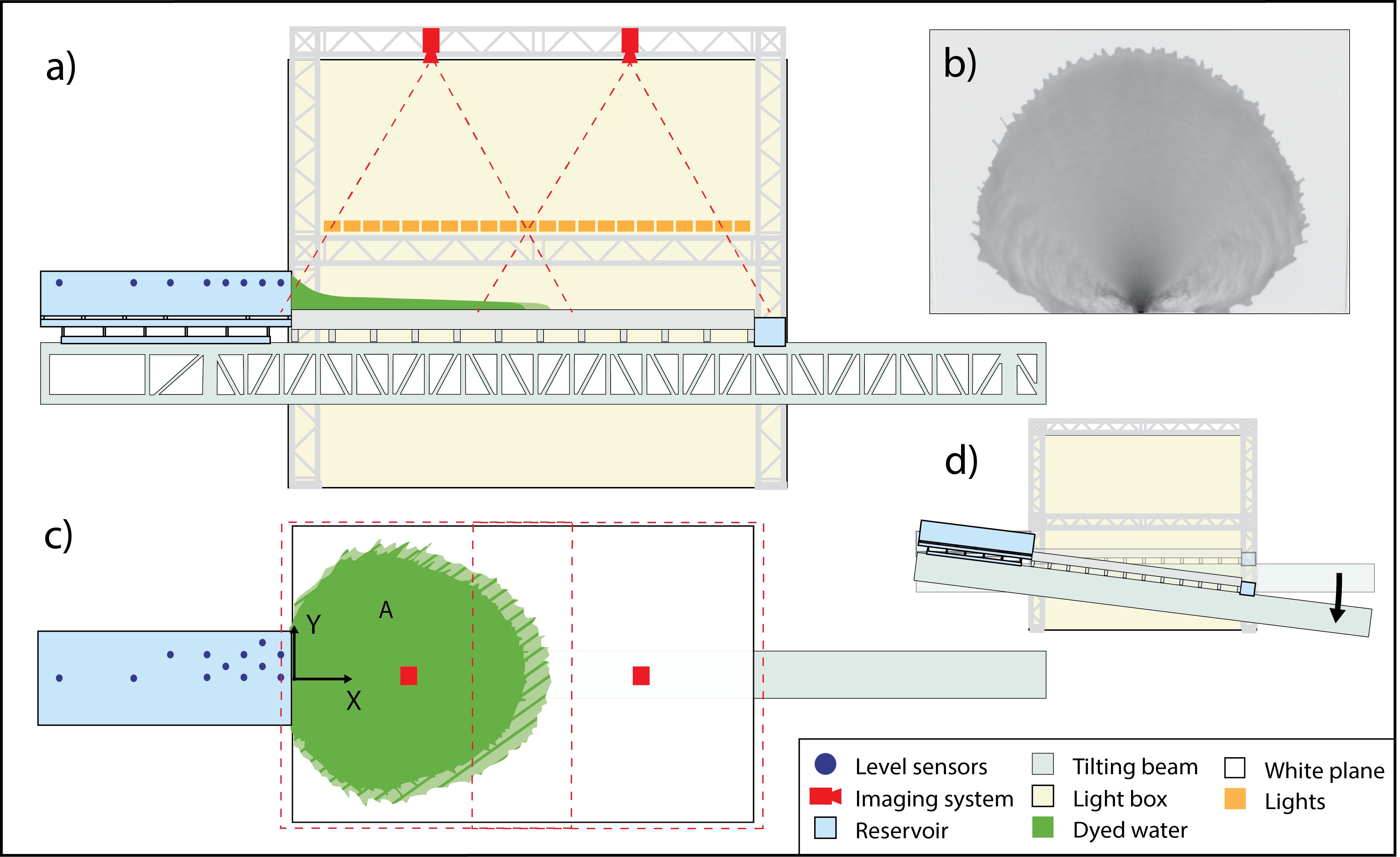}
\caption{A schematic of the instrument setup. a) Side view with the dam released, b) captured dam-break wave with $H_o=0.30 \ m$ at $t=1.6s$,  c) top view with the positions of the measuring instrumentation and reference system and d) illustration of the tilting kinematic.}
\label{Fig:Setup_2D_2}  
\end{figure*}

Wave-depth measurements over the plane are obtained from the two scientific cameras mounted at the top of the light box (Figure ~\ref{Fig:Setup_2D_2}). The cameras are Andor Zyla 
CMOS ($5.5 \, \mathrm{MP}$) (Oxford Instruments, Abingdon, UK), acquiring $16$-bit grayscale images at a resolution of $2560 \times 2160$ pixels, with a ground pixel size of $1.66 \, \mathrm{mm}$ at the level of the plane floor. Cameras mount SAMYANG lenses model MF 14mm F2.8 MK2 (LK Sanyang, Masan, South Korea).  These images form the basis for reconstructing the depth field $h(x,y,t)$, as described in the following section. 

To provide an independent validation of the image-based depth measurements, the wave volume is also evaluated through a continuity-based approach: at each time instant, the integrated water volume obtained from $h(x,y,t)$ must equal the volume lost from the upstream reservoir. However, accurately determining the latter is non-trivial because the free surface inside the reservoir is highly disturbed after breach opening, with an upstream-traveling surge \cite{zweifel2006plane,heller2011wave} and pronounced transverse oscillations. 

These complex surface dynamics prevent a reliable level estimate from a single gauge. For this reason, twelve ultrasonic level sensors (Balloff, model BUS004W) were installed in the upstream reservoir. Their spatial distribution was optimised through several iterations to obtain a robust and stable estimate of the reservoir volume over time, despite the presence of longitudinal and transverse waves.

Camera coordinates in the breach-centered reference system (Figure \ref{Fig:Setup_2D_2}) are $X_c= 2.06 \ m$ $Y_c= 0.12 \ m$ $Z_c=4.7 \ m$, level sensors coordinates are reported in Table \ref{tab:sens_coordinates}. 

Data acquisition is managed through a National Instruments interface card, which controls both the ultrasonic level sensors and the camera trigger. LabVIEW (National Instruments, Austin, Texas, USA) is used to operate the breach-opening mechanism and to record the sensors' signals, while the image acquisition is handled by Andor’s proprietary software. Images was analized with MATLAB (MathWorks, Natick, Massachusetts, USA) to pass from pixel to metric coordinates as described elsewhere \cite{melis2019resistance}. Wave depth is then quantified by correlating the grayscale levels of the recorded pixels with known water depths, adapting the established transmitted-light methodology of \citet{aureli2011image}. This technique, central to the present study, is described in detail in the following section. 

\begin{table}[h]
\caption{Ultrasonic level sensors locations}\label{tab:sens_coordinates}
\begin{tabular*}{\textwidth}{lcccccccccccc}
& $S_1$ & $S_2$ & $S_3$ & $S_4$ & $S_5$ & $S_6$ & $S_7$ & $S_8$ & $S_9$ & $S_{10}$ & $S_{11}$ & $S_{12}$ \\
\midrule
$-x \ [\mathrm{mm}]$ & 100 & 100 & 316.5 & 316.5 & 533 & 533 & 749.5 & 966 & 966 & 1399 & 1832 & 2698 \\
$y \ [\mathrm{mm}]$ & 0   & 250 & 125   & 375   & 0   & 250 & 125   & 0   & 250 & 250  & 0 & 0    \\
\botrule
\end{tabular*}
\footnotetext{Note: the $z$ coordinate is adjustable to be kept barely above the initial free surface.}
\end{table}

\section{Imaged water depth}
The methodology for measuring water depth relies on the principle of light attenuation, described by the  Beer–Lambert law \cite{beer1852bestimmung}. For a given wavelength $\lambda$, this law states that the transmitted light intensity decays exponentially along an attenuation path due to absorption and scattering:
\begin{equation}
\label{eq:beer-lambert_law}
i/i_o = \exp\left(-a l_p\right),
\end{equation}
where $i$ and $i_o$ are, respectively, the transmitted and incident intensities, $a$ is the attenuation coefficient, and $l_p$ is the optical path length through the medium. 

Since the light emitted by the LEDs and detected by the camera spans a broad wavelength range, effective attenuation cannot be described by the monochromatic form of the Beer–Lambert law alone. Its determination involves three spectral components: i) the emission spectrum of the LED illumination $i_o(\lambda)$, ii) the wavelength-dependent absorption of the dye $a(\lambda)$ and iii) the spectral sensitivity of the camera sensor $QE(\lambda)$ (also known as quantum efficiency). In this framework, the bulk attenuation refers to the physical attenuation experienced by the incident light as it travels through the colored water layer, resulting from the spectral integration of the Beer–Lambert law on the LED emission and dye absorption spectra. In contrast, effective attenuation is the quantity actually retrieved from camera images: it corresponds to the bulk-attenuated light after being filtered by the camera’s spectral sensitivity and mapped into grayscale levels. This is the signal that was later correlated with the depth of the water.

\begin{figure}[ht!]
\centering
\noindent\includegraphics[width=31pc]{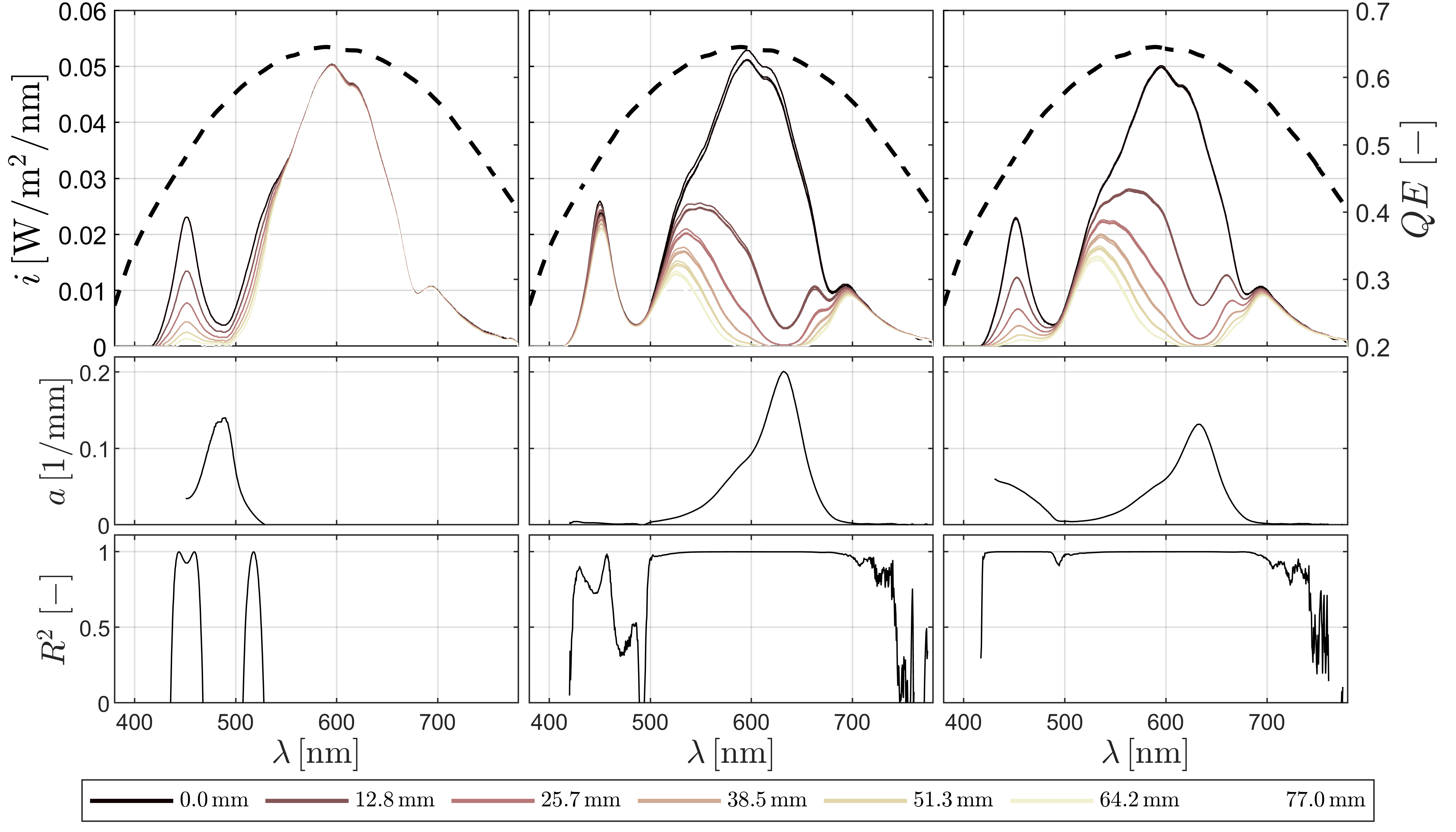}
\caption{Results of spectral investigation on light attenuation for three food grade dyes as function of light wavelength $\lambda$. On column from left to right: yellow, blue and green dyes. On rows from top to bottom: spectral light intensity $i$ as $\mathrm{W/m^2/nm}$ (left axis) for various attenuation path lengths $l_p$ (color lines) and CMOS camera sensors' sensitivity $QE$ (dashed lines, values on right axis), best fit attenuation coefficient $a(\lambda)$ from Eq. \ref{eq:beer-lambert_law}, coefficient of determination $R^2(\lambda)$ of the fit.}
\label{Fig:spectra}  
\end{figure}

The first step of the calibration therefore consists in characterizing the light-source spectrum and how the dye filters the incident light at each wavelength (bulk transmitted intensity).  The bulk transmitted intensity  $I$ and its normalized form  $I/I_o$ follow from integrating the Beer–Lambert law \ref{eq:beer-lambert_law} relation across the full spectrum of the light:
\begin{equation}
\label{eq:bulk_beer}
I/I_o =\frac{1}{I_o} \int i(\lambda) \exp\left(-a(\lambda) l_p\right) d\lambda,
\end{equation}
where $I_o$ is the bulk incident light. Four food grade color dyes were tested  to determine their light-attenuation performance: red ($10\%$ Ponceau 4R, $90\%$ sodium sulfate), yellow ($10\%$ Tartrazine, $90\%$ sodium sulfate), blue ($10\%$ Patent Blue V, $90\%$ sodium sulfate) and green ($5\%$ Tartrazine, $5\%$ Patent Blue V, $90\%$ sodium sulfate). The test were conducted in a cylindrical pipe, sealed with a transparent PMMA bottom filled with colored water at varying depths. Illumination was provided from below using the same LED source adopted in the dam-break experiments, and the transmitted light was recorded from above. To minimize reflections and diffuse scattering, the inner surface of the tube was painted matte black. Light attenuation across the visible spectrum  ($\lambda=350-800 \ nm$) was measured using a Sekonic C-700 SpectroMaster spectrometer. Figure \ref{Fig:spectra} reports the visible light spectra for various water depth (top panels), the corresponding attenuation coefficient $a(\lambda)$ obtained from Eq. \ref{eq:beer-lambert_law} (central panels), and the coefficient of determination $R^2$ associated with a monochromatic exponential fit (bottom panels) for yellow, blue, and green dye solutions prepared at a concentration of $0.5 \ g/l$.

Zero depth light spectra for three tests on figure \ref{Fig:spectra} shows typical white light-emitting diodes' (LEDs) bimodal spectral power distribution consisting of a narrow, blue emission peak centered at approximately $450–465 \, \mathrm{nm}$, superimposed on a broad, yellow-green-to-red band ($500–700 \, \mathrm{nm}$). The yellow dye show's prominent light attenuation in the blue peak and low to no effect on the yellow-green-to-red band; the blue dye complements the warm side of the color spectra with substantially no effect on the blue peak. Generally speaking each dye it's "transparent" to his own color, so has expected the green dye resulting form an even mixing of the previous two dyes shows a good attenuation in both two bands of the bimodal distribution resulting as the must effective in bulk light attenuation. In all three tests, where a consistent attenuation is observed, perfect agreement ($R^2\sim1$) with Eq. \ref{eq:beer-lambert_law} is also observed (red dye attenuation is not reported as substantially similar to yellow dye's one). 

The analysis above characterizes the illumination spectrum of the LED source and the spectral filtering induced by the dyes, thereby defining the bulk attenuation of the incident light. By superimposing the spectral sensitivity of the camera sensor, and recalling that our objective is to maximize the attenuation observed in the effective transmitted intensity, we note that the blue dye produces the strongest reduction of the detectable signal for a given depth. Although the yellow dye also suppresses part of the illumination spectrum, it does so mainly in wavelength regions where the sensor responsivity is low. An effective attenuation model must therefore account for attenuation occurring specifically in the spectral bands where the sensor is most sensitive, and must relate this attenuated signal to the grayscale levels recorded by the camera. 

To relate the bulk attenuation to the signal measured by the camera, we assume that the recorded gray level $G$ is proportional to the effective transmitted intensity reaching the sensor. The normalized grayscale value 
$g_n=G/G_o \sim I/I_o$ where $G_o$ is the gray level corresponding to the incident intensity  $I_o$ after being filtered by the spectral sensitivity of the sensor, will hereafter be referred to as the background gray level. 

\begin{figure}[ht!]
\centering
\noindent\includegraphics[width=31pc]{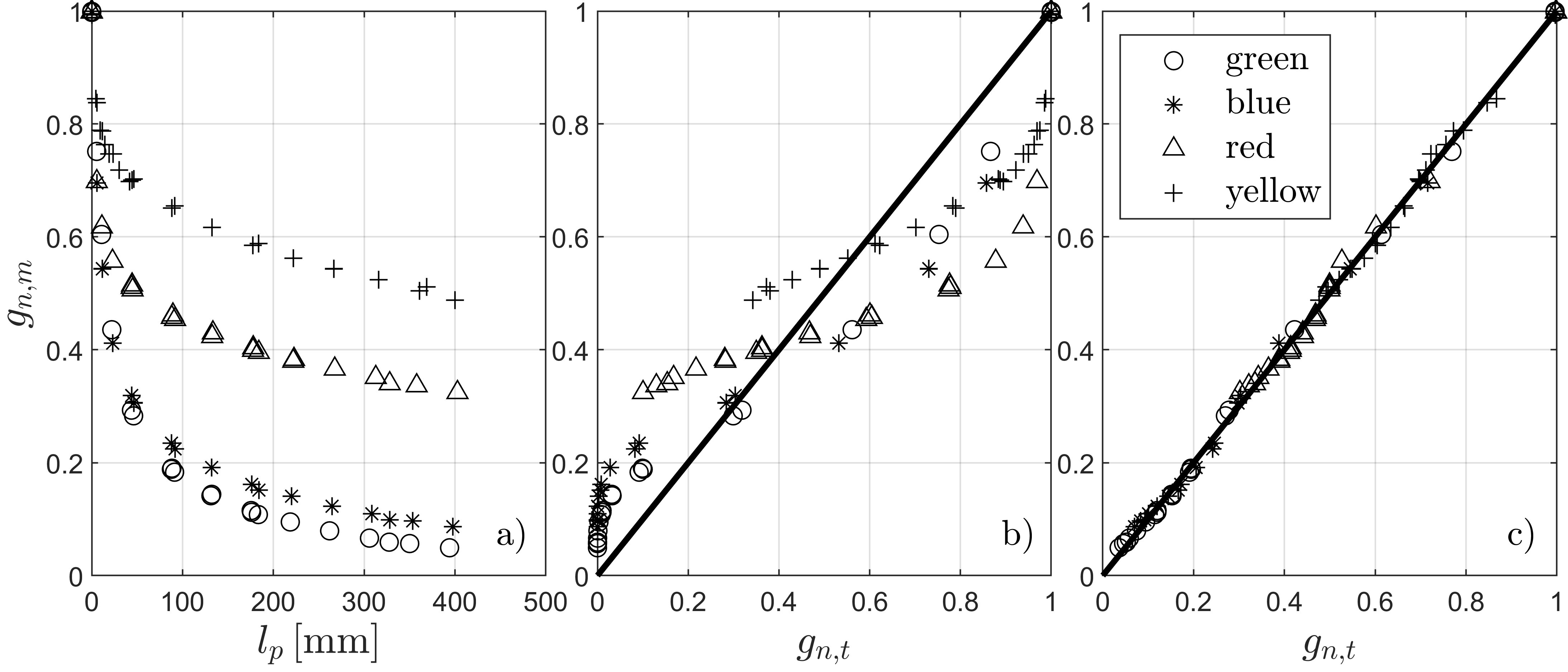}
\caption{On panel a) imaged normalized gray level $g_{n,m}$ along with attenuation path length $l_p$ for four food grade dye (green, blue, red, yellow). On panel b) $g_{n,m}$ along with the same quantity $g_{n,t}$ modeled with Beer Lambert law Eq. \ref{eq:beer-lambert_law}. On Panel c) measured $g_{n,m}$ and modeled $g_{n,t}$ with here proposed bi-exponential model \ref{eq:h_gray_tfun}. On panels b) and c) solid black lines represent perfect agreement with models.}
\label{Fig:attenuation_test}  
\end{figure}

The normalized transmission $g_n$ was then measured experimentally by replacing the spectrometer used in the spectral tests with the same camera employed in the dam-break experiments, while keeping the illumination and optical setup identical. This configuration allowed us to image the bulk-attenuated light under the same conditions adopted in the light-box facility. 
Figure \ref{Fig:attenuation_test}a reports the resulting values of $g_n$ for all four dyes, tested at a concentration of $0.5 \, \mathrm{g/l}$, as a function of the optical path length  $l_p$, All dyes exhibit a similar exponential decay of transmitted intensity with depth, and the green dye confirms the strongest attenuation in accordance with the spectral analysis. 

Because the attenuation coefficient  $a(\lambda)$ varies significantly across the spectrum (see figure \ref{Fig:spectra}), the integral in Eq. \ref{eq:bulk_beer} does not reduce  to a closed-form algebraic solution.  As a consequence, the monochromatic Beer–Lambert law cannot accurately represent the bulk attenuation measured by the camera. This limitation is evident in Figure \ref{Fig:attenuation_test}b, where the normalized gray levels predicted with Eq. \ref{eq:beer-lambert_law} (fitted through nonlinear least-squares regression to obtain $g_{n,t}$) systematically differ from the measured values ($g_{n,m}$). To account for the combined influence of the illumination spectrum, the dye absorption curve and the camera sensitivity, the effective attenuation can be approximated with a sum of exponential \cite{aureli2014combined}. In this case we choose a bi-exponential model in the form:   
\begin{equation}
\label{eq:h_gray_tfun}
        g_n = C_1\exp(-c_1 l_p) + C_2 \exp(-c_2 l_p),
\end{equation}
where $C_1$ and $C_2$ are amplitude coefficients and $c_1$ and $c_2$ are attenuation constants. As shown in Figure  \ref{Fig:attenuation_test}c the bi-exponential model fitted through nonlinear least-squares regression provides an excellent representation of the measured attenuation curve, with a regression error of $1-R^2=5.9\text{x}10^{-4}$. 

The bi-exponential model above provides a compact representation of the effective attenuation under controlled optical conditions. However, its use in the dam-break experiments requires an additional calibration step carried out directly within the light-box setup. In this environment, the illumination field, the camera geometry and the reflective properties of the floor and walls differ from those of the column tests, and the depth–intensity relationship must therefore be determined in situ. For this purpose, a dedicated calibration campaign was conducted on the experimental floor, as described below.  

\begin{figure}[ht!]
\centering
\noindent\includegraphics[width=31pc]{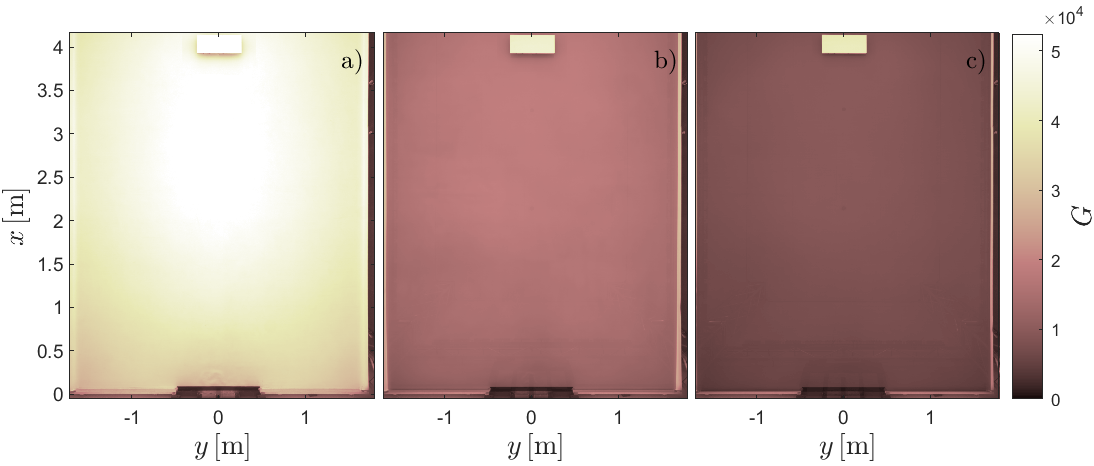}
\caption{Examples of imaged gray level field $G$. On panel a) background gray level, on panels b) and c) two water depths $h_v= 13.7, \, 44.6 \, \mathrm{mm}$ respectively. The reference patch is visible in images' top corner.}
\label{Fig:bat_cal_pics}  
\end{figure}

For calibration  purposes, the parameters of Eq. \ref{eq:h_gray_tfun} were determined using 33 standing-water levels ranging from $1.3 \,  \mathrm{mm}$ to $44.6 \, \mathrm{mm}$, imposed on the plane floor. To ensure significant decay of the normalized grayscale value $g_n$ over the expected depth range, a concentration of $0.172 \, \mathrm{g/l}$ of green dye was added to the water. The water level over the plane was measured with a vernier scale connected to a piezometric pipe installed in the upstream tank, and one image was acquired for each water level. To obtain a reliable map of the normalized grayscale values $g_n$ the images were processed through a two-step normalization procedure. First, each image was normalized with respect to a background image of the dry floor, this compensate the spatial variability of the of the incident light. Second, a further normalization was applied using the mean gray level of a reference patch kept permanently above water. Although this second step is unnecessary for backlit bottom-surface imaging, it's essential here because increasing water coverage reduces the incident light intensity, since the floor is a major secondary light source (up to 15\% at $h_v=33 \, \mathrm{mm}$). In Figure ~\ref{Fig:bat_cal_pics}, the background image of the floor with the reference patch is shown (panel a) as well as an example of two captured water depth of $h_v=13.7 \, \mathrm{mm} \, 44.6 \, \mathrm{mm}$ on panels b and c respectively.

\begin{figure}[ht!]
\centering
\noindent\includegraphics[width=15pc]{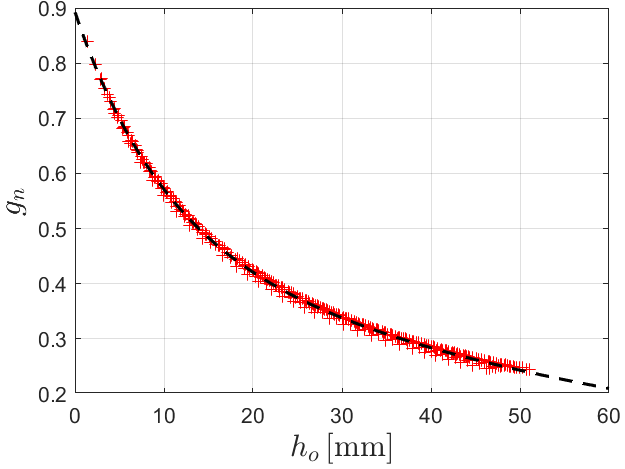}
\caption{The computed optical depths $h_o$ vs normalized gray level $g_n$ (red cross) along with equation \ref{eq:h_gray_tfun} parametrized by least square regression (dashed black line). Best parameters are $C_1=0.38$, $c_1=0.12$, $C_2=0.51$ and $c_2=0.016$ with dye concentration of $172 \ g/m^3$.}
\label{Fig:bat_cal_depth_correction}  
\end{figure}

The level read on the piezometric pipe $h_v$ has been corrected to take into account the effective optical path in the liquid. Assuming that the incidence/transmitted light is not perfectly normal to the free surface, the absorption follows Beer-Lambert's law as a function of the path length in water $l_p= 2h_v/\cos(\theta_w)$, with \(\theta_w\) obtained from Snell's law $n_a\sin\theta_a = n_w\sin\theta_w$ where $n_a=1$ and $n_w=1.33$ are well known refractive indexes for air and water restrictively and $\theta_a$ is the angle of light with the free surface (on air side) as can be easily retrieved from camera position and orientation in respect of the floor plane. Let us denote by $h_o(x,y)$ the optically corrected level used in subsequent analyses ($h_o=l_p/2$). Figure \ref{Fig:bat_cal_depth_correction} shows the computed $h_o$ against the measured $g_n$ for a subsample of points evenly distributed over the floor (red cross) along with Eq. \ref{eq:h_gray_tfun} parametrized trough nonlinear least square regression (black dashed line). The calibration points exhibit excellent agreement with the here proposed bi-exponential model with a regression error of $1-R^2=1.1\text{x}10^{-3}$ confirming its applicability in the light box configuration.  

\section{Estimation of the discharge through the breach}

A key element of the system is the discharge through a relatively narrow breach, which controls both the initial outflow rate and the subsequent wave dynamics. Under inviscid free-flow conditions at the breach, the volumetric flow rate $Q$ can be estimated by applying an energy conservation argument based on Bernoulli’s equation \cite{bernoulli1738hydrodynamica}. The discharge velocity is $u_c(z) = \sqrt{2g\,(H - z)}$ where $z$ is the vertical coordinate measured from the breach bottom. Including a contraction coefficient $C_c$ to account for \textit{vena contracta} effects, the total discharge becomes:

\begin{equation}
    Q(H) 
    = (b\,C_c) \int_{0}^{H} \sqrt{2g\,(H - z)}\,dz
    = \frac{2}{3}\,b\,C_c\,\sqrt{2g}\,H^{3/2},
\end{equation}

where $H$ is the upstream water depth far from the breach. The temporal evolution of $H$ follows from the continuity equation $dH/dt = -Q(H)/A(H)$, where $A$ is the planform area of the reservoir. Solving this ordinary differential equation yields to explicit expressions for $H(t)$ and $Q(t)$:

\begin{equation}
\label{eq:H of time}
    H(t)
    = H_o 
      \Bigl(
        1 + \frac{b\,C_c}{3\,A}\,t/T_o
      \Bigr)^{-2},
\end{equation}

\begin{equation}
\label{eq:Q of time}
    Q(t)
    = Q_o 
      \Bigl(
        1 + \frac{b\,C_c}{3\,A}\,t/T_o
      \Bigr)^{-3},
\end{equation}

where $Q_o = (2/3)\,b\,C_c\,\sqrt{2g}\,H_o^{3/2}$ is the initial discharge and $T_o=1/\sqrt{2gH_o}$. The cumulative volume released from the tank, obtained as the complement of the in-tank volume $V_t=HA$, becomes:
\begin{equation}
\label{eq:V of time}
    V(t)
    = V_o\left[1- 
      \Bigl(
        1 + \frac{b\,C_c}{3\,A}\,t/T_o
      \Bigr)^{-2}\right],
\end{equation}
where $V_o$ is the initial water volume. This simplified frictionless model neglects the complex emptying dynamics induced by negative wave propagation and reflection, and does not account for finite velocity far from the breach. Nevertheless, it provides a reliable upstream boundary condition for both one-dimensional and radially spreading dam-break flows when $bHC_c<<A$, a condition that is fully satisfied in the present experiments and is also typical of real dam-break scenarios, where the breach width is small compared with the reservoir surface area. Because the discharge depends directly on the upstream water depth, an accurate measurement of the reservoir level is essential for validating the theoretical emptying model and for calibrating the discharge coefficient $C_c$.

\begin{figure*}[ht!]
\centering
\noindent\includegraphics[width=31pc]{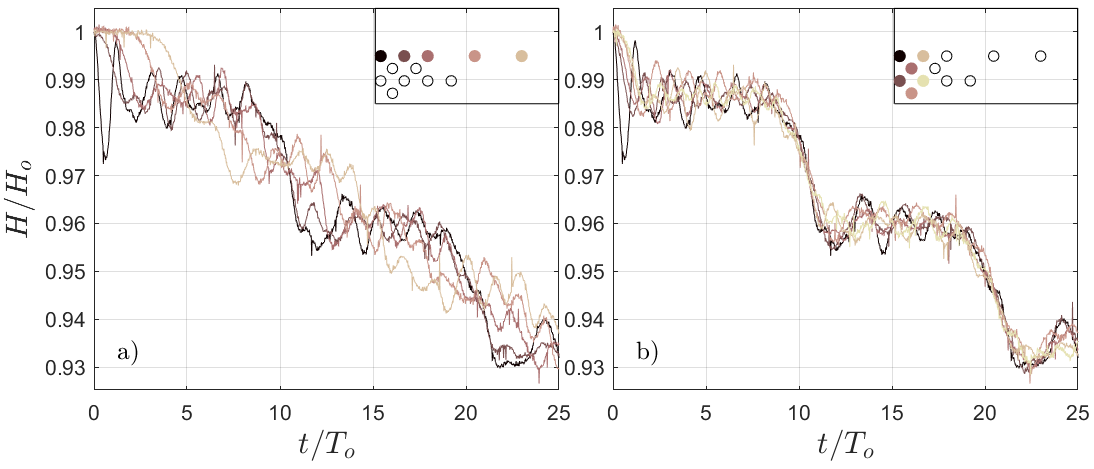}
\caption{Example of recordings from the ultrasonic sensors during tank emptying (normalized level \(H/H_0\) vs. normalized time \(t/T_0\)). On top left corner  on each panels the location of the sensors inside the reservoir. Panel a) shows level sensors aligned with the reservoirs' symmetry axis, panel b) shows the six sensors closest to the breach. The stepwise trend reflects the arrival of the negative surge at the various measurement points.}
\label{Fig:sensors}  
\end{figure*}

To record the level variations generated by the negative surge induced of the gate, twelve ultrasonic sensors were installed inside the upstream reservoir in a horizontal array positioned above the maximum expected water level. A representative example is shown in Figure ~\ref{Fig:sensors}. The plots of the normalized water level, $h/H_0$, as a function of the non-dimensional time $t/T_0$, show a stepped pattern, caused by the sequential arrival of the negative surge at the different measurement locations. Each discontinuity corresponds to the passage of a lowering front across the tank, while the level remains nearly constant between successive fronts, producing the characteristic stepwise deepening. 

\begin{figure*}[ht!]
\centering
\noindent\includegraphics[width=15pc]{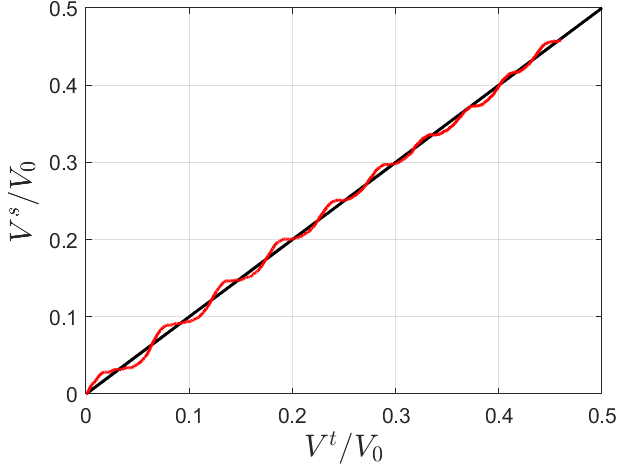}
\caption{Theoretical normalized wave's volume $V^t/V_o$ from Eq. \ref{eq:V of time} along with measured with level sensors $V^{s}/V_o$ (red line). Solid black line represent perfect agreement with theory.}
\label{Fig:volume_comp}  
\end{figure*}

The total volume of the dam-break wave is independently quantified using the ultrasonic measurements. The sensor-based volume $V^s$ is obtained by weighting each depth measurement according to its associated Voronoi area partition  \cite{fortune2017voronoi}. Figure \ref{Fig:volume_comp} compares the normalized sensor-derived volume \(V^{t}/V_0\) with the theoretical equation in ~\ref{eq:V of time} using a contraction coefficient $C_c=0.65$ obtained via nonlinear regression. The agreement is very good, apart from small oscillations. These fluctuations arise from the nonuniform spatial distribution of the sensors, which causes localized biases when the negative surge crosses the tank. The oscillation period of $V/V_o\sim0.063$ is consistent with the surge round-trip time $2L_t/\sqrt{gHo}$, for which Eq. ~\ref{eq:V of time} yields $V/Vo=0.061$.

\section{Results}
In this section, we present and discuss our preliminary results. We performed 15 experiments, divided into three initial heights ($H_o=0.20; \ 0.25; \ 0.30 \, \mathrm{m}; $) in the reservoir, each repeated 5 times to assess statistical repeatability. For each test, the imaged gray level were metrically rectified and resampled into uniform  \(g_n(x,y,t)\). The depth level field was then i) aligned temporally on the opening trigger ii) masked to remove artifacts near the edges. The $h_o$ is than computed form $g_n$ with the parametrized attenuation curve (Eq. \ref{eq:h_gray_tfun}) and finally $h(x,y,t)$ is retrieved under the simplified assumption of horizontal free surface (generalized slope-aware free surface reconstruction procedure on Appendix A). For each initial height $H_o$, the ensemble average depth map $\langle h \rangle(x,y,t)$ and standard deviation map $\sigma_h(x,y,t)$ has been computed for the 5 runs. 

\begin{figure*}[ht!]
\centering
\noindent\includegraphics[width=31pc]{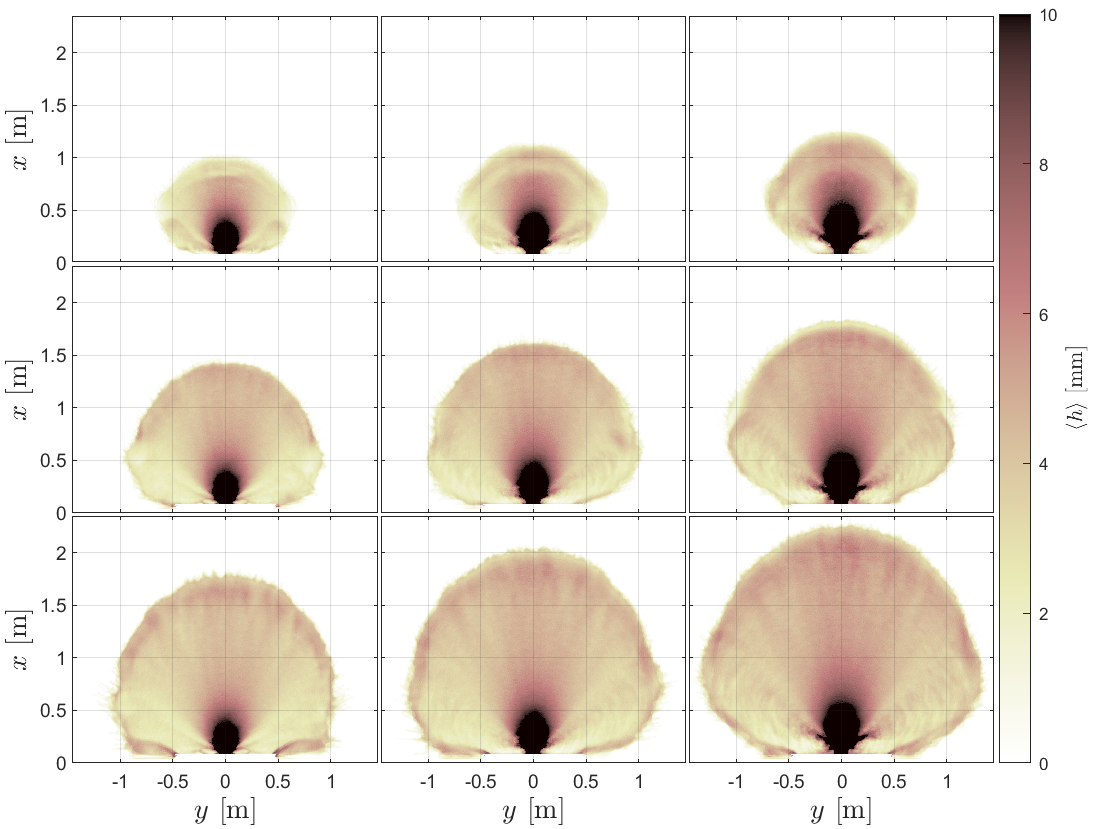}
\caption{Ensemble averaged depth map $\langle h \rangle(x,y)$ for all experimental conditions at three different time instants. On columns, from the left to the right, $H_o=0.20; 0.25; 0.30 \, \mathrm{m}$. On rows, from the top to the bottom, instants $t=0.63; 1.13; 1.63 \,  \mathrm{s}$.}
\label{Fig:h_map}  
\end{figure*}

Figure \ref{Fig:h_map} shows the results at three representative instants ($t= 0.63 ; \ 1.13; \ 1.63 \ s $) common to all $H_o$. The figure is organized in a 3 column (one for each $H_o$) and 3 rows (for time instants) array; $\langle h \rangle (x,y,t)$  is represented as colour scale (common for all panels). The depth fields show that the front propagation velocity increases almost linearly with the initial reservoir level $H_o$. In all cases the free surface is very steep in the near-gate region, where $h$ drops within about $0.5 \, \mathrm{m}$; farther downstream the depth continues to decrease but with a significantly lower surface slope. Within a central sector of roughly $90°$ around the x-axis the elevation field remains remarkably smooth, whereas at larger angles slight surface instabilities it's observed. Small phase shifts between repetitions smear out the pattern of those instabilities in the ensemble average, therefore them amplitude and frequency must be analysed on individual realizations. Furthermore, it is not meaningful to use these lateral regions to assess precision of the depth reconstruction.

\begin{figure*}[ht!]
\centering
\noindent\includegraphics[width=31pc]{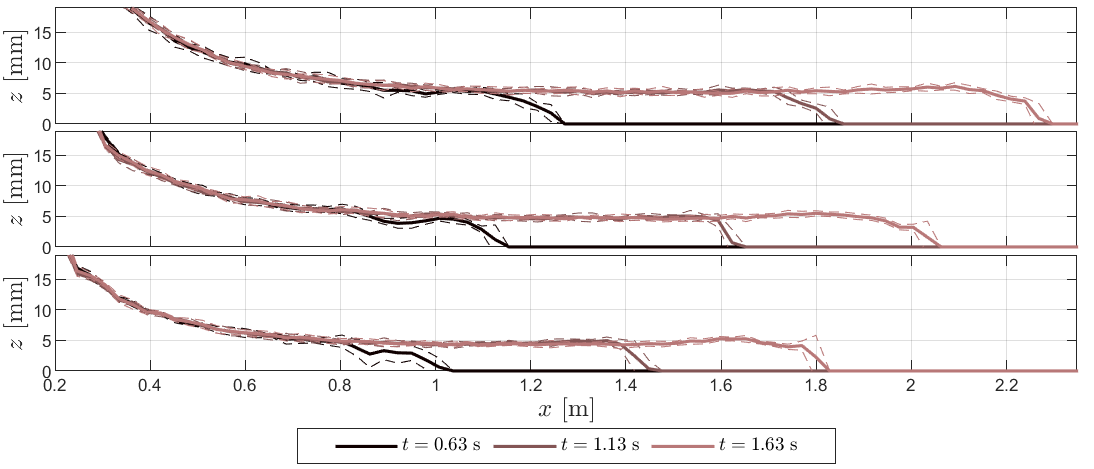}
\caption{Free surface profiles from vertical section aligned with $x$ axis, for three time instants $t=0.63; 1.13; 1.63 \,  \mathrm{s}$  (colour line). On panels, from the bottom to the top, $H_o=0.20; 0.25; 0.30 \, \mathrm{m}$. Solid lines represent ensemble averaged depth $\langle h \rangle$ while dashed lined represent 95\% confidence bounds computed among 5 runs.}
\label{Fig:h_sections}  
\end{figure*}

 To better understand the evolution of $\langle h \rangle$, vertical sections of the free surface aligned with the $x$ axis were computed .  Figure \ref{Fig:h_sections}  shows $\langle h \rangle(x,0)$ (solid lines) for $H_o=0.20; 0.25; 0.30 \, \mathrm{m}$ on panels from the bottom to top and for the same three time instants of Figure \ref{Fig:h_map}  (as colour line); 95\% confidence bounds, computed among the 5 runs, are also reported as dashed lines. As expected,  the vertical sections show how $\langle h \rangle$  scales roughly with $H_o$.  Also, the free surfaces from different time instants substantially overlap, except for a short region close to the advancing front, this behaviour is observed consistently among initial water depths $H_o$  and it is in line with the small breach condition (i.e. $bH_oC_c<<A$) where the minor loss of reservoir's volume over time leads to a quasi- steady boundary condition. Finally, those profiles further highlights how the surface slope drastically decreases shortly after the breach , with $-\partial h/\partial x<0.05$  for $z>0.2 \ \mathrm{m}$ and $\partial h/\partial x\sim 0$ for $z>0.7 \ \mathrm{m}$; this mostly sub-horizontal surface justify the assumption of locally horizontal free surface on $h$ reconstruction .

\begin{figure*}[ht!]
\centering
\noindent\includegraphics[width=31pc]{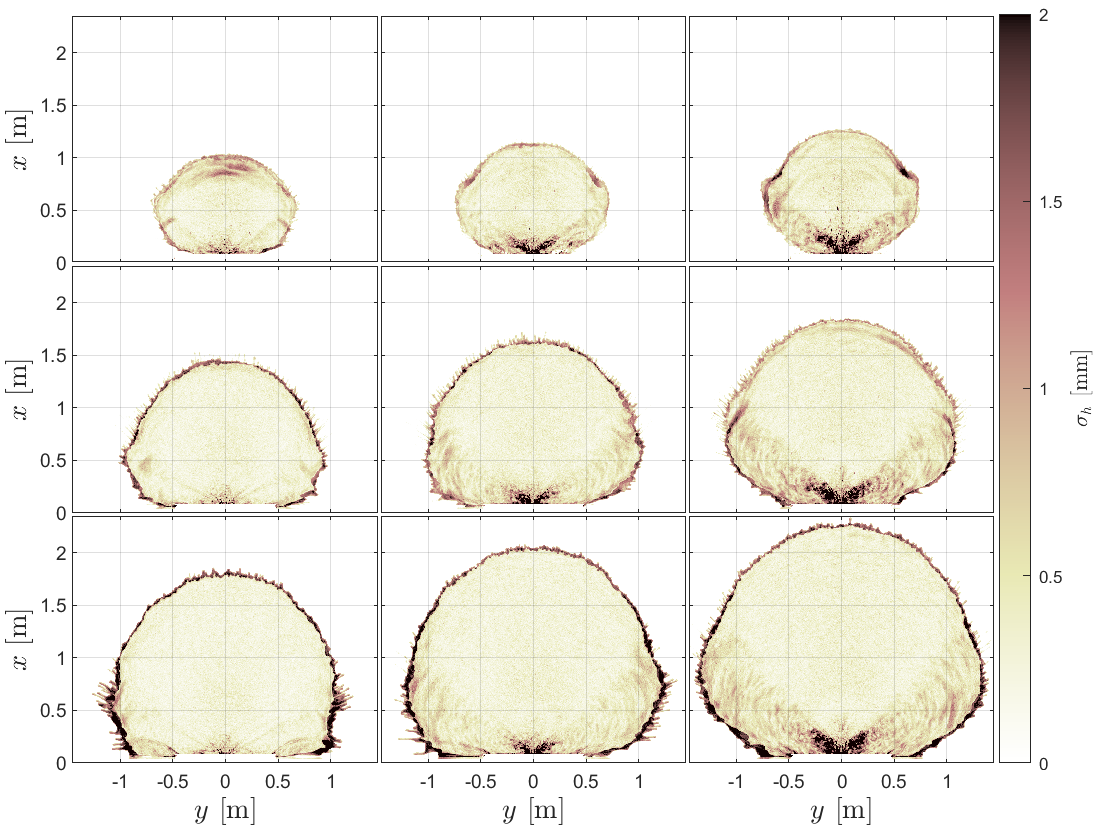}
\caption{Depth standard deviation map $\sigma_h(x,y)$ for all experimental condition at three different time instants. On columns, from the left to the right, $H_o=0.20; 0.25; 0.30 \, \mathrm{m}$. On rows, from the tom to the bottom, instants $t=0.63; 1.13; 1.63 \,  \mathrm{s}$.}
\label{Fig:cv_map}  
\end{figure*}

The same layout adopted for represent $\langle h \rangle$ has been used in Figure\ref{Fig:cv_map} for $\sigma_h$ fields. Typically the largest $\sigma_h$ values are confined to a thin annulus that follows the advancing front, as expected given the not perfect superposition of wave fronts among repetition. Although, the interior of the lobe exhibits much smaller run-to-run variability with $\sigma_h \sim 0.3 \ \mathrm{mm}$. For higher $H_o$, in the central sector of about $90°$ around the x-axis, where the free surface remains relatively smooth, $\sigma_h$ stays low for all $H_o$ and times, indicating a good repeatability of the imaged depth. Laterally, in the regions where short-crested waves were observed in the mean elevation maps, the standard deviation increases. In these regions $\sigma_h$ is mainly controlled by slight phase changes of surface instability, rather than by the random error of the gray depth conversion. Because such rapidly evolving instabilities cannot be synchronized across realizations, the large $\sigma_h$ values along the front and at the sides should not be interpreted as a local loss of accuracy of the measurement technique, but rather as a limitation of ensemble averaging.

\begin{figure*}[ht!]
\centering
\noindent\includegraphics[width=15pc]{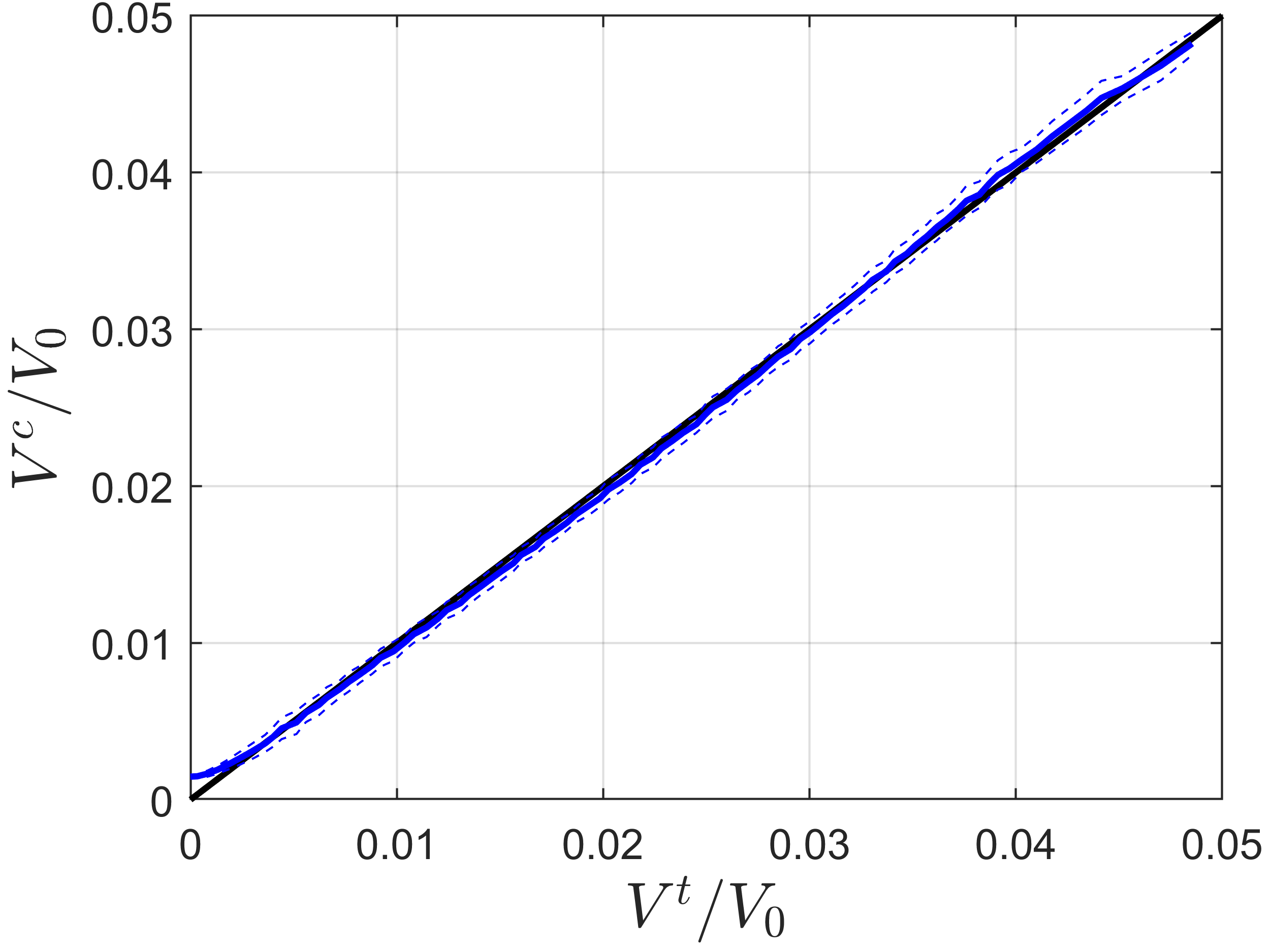}
\caption{Theoretical normalized wave's volume $V^t/V_o$ from Eq. \ref{eq:V of time} along with ensemble averaged imaged wave's volume $V^{c}/V_o$ (solid blue line), dashed blue lines represent 95\% confidence bounds computed among 15 runs. Solid black line represent perfect agreement with theory.}
\label{Fig:volume_comp_camera}  
\end{figure*}

Finally, the total imaged volume of the dam-break wave  $V^c$ is obtained via surface integral of the imaged $h$. Additionally, a correction to $V^c$ is applied to account for the volume in a blind zone adjacent to the dam wall, where water depths could not be directly retrieved due to reflection artifacts. The blind zone extends crosswise as the tank width and 0.1 m streamwise. This correction assumes an outflow vein of height $H$, width $bC_c$, and a 45° slope \cite{bernoulli1738hydrodynamica}, while regions outside this vein are assigned the average depth measured on the plan. The contribution of this correction remains minimal, approximately $0.1\%$ of $V_o$ an substantially constant in time. Figure \ref{Fig:volume_comp_camera} compares the imaged volume \(V^{c}/V_0\) with the theoretical law Eq. \ref{eq:V of time} with $C_c=0.65$ calibrated trough non-linear regression of measured $V^s$. This results highlight two key aspects: i) the agreement between $V^c$ and the theoretical estimate confirms the reliability of the imaging-derived water depths and ii) the in-tank volume and average in-tank depth vary by less than $5\%$, supporting the assumption of negligible variation of in-tank water level on time $H \sim H_o$. 

\section{Conclusion}
In this work we introduced a new large-scale experimental facility designed for the investigation of fully three-dimensional dam-break flows spreading over a wide, effectively unconfined plane, together with an image-based technique for reconstructing the corresponding space--time evolution of water depth. The key  enabling element of the facility is the light box, which produces controlled and highly uniform diffuse illumination over a $6.4\,\mathrm{m}\times 3.4\,\mathrm{m}$ plane through an array of sixty LED sources, while avoiding direct lighting of the free surface and therefore minimizing specular reflections. Combined with overhead scientific CMOS cameras, this configuration ensures  repeatable imaging conditions over an area that would be impractical to illuminate by conventional backlighting approaches. 

The proposed depth-reconstruction methodology is based on transmitted-light attenuation and explicitly accounts for the broadband nature of the illumination and for the spectral sensitivity of the camera sensor. A preliminary spectral characterization of the LED emission and dye absorption was used to identify a dye mixture providing strong attenuation within the effective sensitivity band of the sensor, thereby maximizing sensitivity to small optical path lengths. Because the resulting attenuation cannot be accurately described  by a monochromatic Beer--Lambert law, a bi-exponential relationship between normalized gray level and optical path length was introduced, yielding an excellent representation of the measured transmission curve. An in situ calibration campaign carried out directly on the experimental plane supported by a two-step normalization procedure and an optical-path correction based on refraction, confirmed the robustness of the bi-exponential model under the specific radiative environment of the light box. 

The methodology was applied to a preliminary campaign of fifteen dam-break experiments with three initial reservoir levels, each repeated five times to assess statistical repeatability. The reconstructed depth fields $h(x,y,t)$ show consistent propagation patterns and a clear dependence of front kinematics on the initial reservoir level. Ensemble variability maps indicate that the largest standard deviations are mainly confined to the advancing front, where small run-to-run shifts are unavoidable, while the interior of the lobe, particularly within a central sector aligned with the main direction of propagation, exhibits very low variability. This behaviour supports the repeatability of the image-based depth reconstruction under rapidly varying three-dimensional conditions. 

A further validation of the reconstructed depth fields was obtained by comparing the total imaged wave volume with independent estimates of the reservoir volume loss. The sensor-based volume derived from the array of ultrasonic level sensors, together with a simplified analytical emptying model calibrated through a discharge coefficient, provides a continuity-based benchmark that is independent of the imaging procedure. The good agreement observed between these independent estimates and the camera-derived volumes confirms the reliability of the proposed methodology and supports its use as a quantitative reference for the validation of numerical models of unconfined dam-break spreading. 

The present contribution provides both an experimental platform and a measurement technique that can be extended in several directions. Future work will exploit the flexibility of the facility to investigate the effects of breach geometry, bed inclination and surface treatments, including roughness elements and obstacles. In addition, the limitations associated with blind zones will be addressed by introducing complementary measurement strategies. Beyond dam-break applications, the combination of controlled diffuse illumination and absorption-based imaging offers a practical route for high-resolution and spatially distributed depth measurements in a broader class of rapidly varying shallow free-surface flows over large planar domains.

\section*{Declarations}


\begin{itemize}
\item Funding: The authors acknowledge support from Fondo europeo di sviluppo regionale (FESR) for project Bacini Ecologicamente sostenibili e sicuri, concepiti per l'adattamento ai Cambiamenti ClimAtici (BECCA) in the context of Alpi Latine COoperazione TRAnsfrontaliera (ALCOTRA) and project Nord Ovest Digitale e Sostenibile - Digital innovation toward sustainable mountain (Nodes - 4). 

\item Conflict of interest/Competing interests: The Authors have no conflicts to disclose.

\item Ethics approval and consent to participate: Not applicable.  

\item Consent for publication: All authors consent for publication.

\item Data availability: The data that support the findings of this study are available from the corresponding author upon reasonable request.

\item Materials availability: The material that support the findings of this study are available from the corresponding author upon reasonable request.

\item Code availability: Not applicable

\item Author contribution: EB wrote the main manuscript. DP provided supervision and edited the manuscript. All authors contribute to project conception, design and realization. All author revised the final version of the manuscript. 

\end{itemize}




\begin{appendices}

\section{Depth Model Derivation}\label{secA1}

This appendix formalizes the reconstruction of the water depth from the attenuation path length $l_p$. Let $(x,y,z)$ be a Cartesian reference system and let the bottom surface be the plane $z=0$. The camera center (pinhole) is denoted by $F\in \mathbb{R}^3$.

For any point $P'(x,y)$ in the metrical imaged wave (i.e. the point inferred by back-projecting the corresponding image pixel on the bottom surface in absence of refraction), let be $P$ its effective position on the floor. In general $P'\neq P$ due to light refraction at the free surface. The light ray that impress $P$ on the camera sensor enter the free surface on point $C$ and emerge from it on point $B$. The underwater segments $\overline{CP}$ and $\overline{PB}$ are assumed to be straight and the iteration at the bottom is assumed to be specular reflection.  

Let's be $d_1=(B-F)/\lVert(B-F)\rVert$, $d_2=(P-B)/l_{p2}$ and $d_3=(C-P)/l_{p1}$, the three unit vector describing the direction of the light traveling segments $\overline{CP}$, $\overline{PB}$ and $\overline{BF}$ respectively. Than $l_{p1}=\lVert(C-P)\rVert$ and $l_{p2}=\lVert(P-B)\rVert$ will be attenuation paths before and after reflection on the floor ($l_p=l_{p1}+l_{p2}$).

Let be $n_f$ the known unit normal of the bottom plane and $n_s=N_B/\lVert N_B \rVert$ the unit normal to the free surface at $B$ (where $N_B=\{\partial h/\partial x|_{(B_x,B_y)}; \partial h/\partial y|_{(B_x,B_y)};-1\}$). 

Applying the well known lows for refraction and reflection $d_2$ and $d_3$ can be computed as:

\begin{equation}
\label{eq:d1-d2}
d_2=\eta d_1 - n_s\sqrt{1-\eta^2\left(1-(d_1\cdot n_s)^2\right) }-\eta n_s( d_1 \cdot n_s)
\end{equation}

\begin{equation}
\label{eq:d2-d3}
d_3=d_2-2(d_2 \cdot n_f)n_f
\end{equation}

where $\cdot$ denotes the scalar (dot) product and $\eta$ is the mediums' refraction index ratio, which for air and water is $\eta=1/1.33$. Given that $P'$ $B$ and $F$ are aligned, direction $d_1$ can be also defined as:

\begin{equation}
\label{eq:d1}
d_1=\frac{P'-F}{\lVert P'-F\rVert}
\end{equation}

Assume that, locally around the ray intersection points, the free surface can be approximated as a plane, hence $n_s$ is constant over the local path containing $B$ and $C$. Under this approximation, the depth over $P(x,y)$ i.e. $h(x,h)$ is given by:

\begin{equation}
\label{eq:depth_model}
h=\frac{(n_s\cdot d_2)(n_s \cdot d_3)}{(n_s\cdot e_z)(n_s \cdot d_3-n_s\cdot d_2)}l_p
\end{equation}

where $e_z=\{0; 0; 1\}$ is the vertical unit vector. Note that for plane and free surface both horizontal (i.e $n_f=n_s=e_z$) Eq. \ref{eq:depth_model} simplify to $h=l_pcos(\theta_w)/2$ where $\theta_w$ is the angle between light and free surface on water side.

Generally, the free surface slope is not known \textit{a priori}. this can be solved assuming $n_s=e_z$,  that allow compute $n_s$, $P$, $P'$, $B$ and $C$ from the resulting $h(x,y,e_z)$ and than solve $h$ iteratively. 




\end{appendices}

\bibliography{mybib}

\end{document}